\documentclass[journal]{IEEEtran}%
\pdfoutput=1
\usepackage{amsfonts}
\usepackage{amsmath}
\usepackage{amssymb}
\usepackage{graphicx}
\usepackage{verbatim}%
\providecommand{\U}[1]{\protect\rule{.1in}{.1in}}

\begin{document}

\title{Unified Quantum Convolutional Coding}
\author{Mark~M.~Wilde~and~Todd~A.~Brun\thanks{Mark M. Wilde and Todd A. Brun are with
the Center for Quantum Information Science and Technology and the Communication Sciences Institute of the Ming Hsieh Department of
Electrical Engineering at the University of Southern California, Los Angeles,
California 90089 USA (E-mail: mark.wilde@usc.edu; tbrun@usc.edu).}}
\maketitle

\begin{abstract}
We outline a quantum convolutional coding technique for protecting a stream of
classical bits and qubits. Our goal is to provide a framework for designing
codes that approach the \textquotedblleft grandfather\textquotedblright%
\ capacity of an entanglement-assisted quantum channel for sending classical
and quantum information simultaneously. Our method incorporates several
resources for quantum redundancy:\ fresh ancilla qubits, entangled bits, and
gauge qubits. The use of these diverse resources gives our technique the
benefits of both active and passive quantum error correction. We can encode a
classical-quantum bit stream with periodic quantum gates because our codes
possess a convolutional structure. We end with an example of a
\textquotedblleft grandfather\textquotedblright\ quantum convolutional code
that protects one qubit and one classical bit per frame by encoding them with
one fresh ancilla qubit, one entangled bit, and one gauge qubit per frame. We
explicitly provide the encoding and decoding circuits for this example and
discuss its error-correcting capability.

\end{abstract}


\section{Introduction}

The goal of quantum Shannon theory is to quantify the amount of quantum
communication, classical communication, and entanglement required for various
information processing tasks
\cite{PhysRevA.55.1613,capacity2002shor,ieee2005dev,arx2005dev}. Quantum
teleportation and superdense coding \cite{book2000mikeandike}\ provided the
initial impetus for quantum Shannon theory because these protocols demonstrate
that we can combine entanglement, noiseless quantum communication, and
noiseless classical communication to transmit quantum or classical
information. In practice, the above resources are not noiseless because
quantum systems decohere by interacting with their surrounding environment.
Quantum Shannon theory is a collection of capacity theorems that determine the
ultimate limits for noisy quantum communication channels. Essentially all
quantum protocols have been unified as special cases of a handful of abstract
protocols \cite{arx2005dev}.

The techniques in quantum Shannon theory determine the asymptotic limits for
communication, but these techniques do not produce practical ways of realizing
these limits. This same practical problem exists with classical Shannon theory
because the proofs that the channel capacities for classical communication
are achievable use random coding techniques that are too inefficient in practice
 \cite{bell1948shannon}.

An example of an important capacity theorem from quantum Shannon theory
results from the \textquotedblleft father\textquotedblright\ protocol
\cite{arx2005dev}. The father capacity theorem determines the optimal
trade-off between the rate $E$ of ebits (entangled qubits in the state
$\left\vert \Phi^{+}\right\rangle ^{AB}\equiv(\left\vert 00\right\rangle
^{AB}+\left\vert 11\right\rangle ^{AB})/\sqrt{2}$) and the rate $Q$ of qubits
in entanglement-assisted quantum communication.\footnote{This protocol is the
\textquotedblleft father\textquotedblright\ protocol because it generates many
of the protocols in the family tree of quantum information theory
\cite{arx2005dev}. The nickname \textquotedblleft father\textquotedblright\ is
a useful shorthand for classifying the protocol---there exists a mother,
grandmother, and grandfather protocol as well \cite{prep2008dev}.} These rates
for quantum communication and entanglement consumption (or generation if $E$
is negative) fall within a two-dimensional capacity region. Suppose that a
noisy quantum channel $\mathcal{N}$ connects a sender to a receiver. Let
$\left[  q\rightarrow q\right]  $ denote one qubit of noiseless quantum
communication and let $\left[  qq\right]  $ denote one ebit of entanglement.
The following resource inequality is a statement of the capability of the
father protocol:%
\begin{equation}
\left\langle \mathcal{N}\right\rangle +E\left[  qq\right]  \geq Q\left[
q\rightarrow q\right]  . \label{eq:father}%
\end{equation}
The above resource inequality states that $n$ uses of the noisy quantum
channel $\mathcal{N}$ and $nE$ noiseless ebits are sufficient to communicate
$nQ$ noiseless qubits in the limit of large $n$. The rates $E$ and $Q$ are
related to the noisy channel $\mathcal{N}$ and there is a mutual dependence
between them so that they form a capacity region. The father capacity theorem
gives the optimal limits on the resources, but it does not provide a useful
quantum coding technique for approaching the above limits.

Another important capacity theorem determines the ability of a noisy quantum
channel to send \textquotedblleft classical-quantum\textquotedblright\ states
\cite{cmp2005dev}. Let $\left[  c\rightarrow c\right]  $ denote one classical
bit of noiseless classical communication. The result of the classical-quantum
capacity theorem is also a resource inequality:%
\begin{equation}
\left\langle \mathcal{N}\right\rangle \geq Q\left[  q\rightarrow q\right]
+R\left[  c\rightarrow c\right]  . \label{eq:CQ}%
\end{equation}
The resource inequality states that $n$ uses of the noisy quantum channel
$\mathcal{N}$ are sufficient to communicate $nQ$ noiseless qubits and $nR$
noiseless classical bits in the limit of large $n$. The capacity theorem
associated to the above resource inequality, in some cases, proves that we can
devise clever classical-quantum codes that perform better than time-sharing a
noisy quantum channel $\mathcal{N}$ between purely quantum codes and purely
classical codes.

The \textquotedblleft grandfather\textquotedblright\ capacity theorem
determines the optimal triple trade-off between qubits, ebits, and classical
bits for simultaneous transmission of classical and quantum information using
an entanglement-assisted noisy quantum channel $\mathcal{N}$
\cite{prep2008dev}. The grandfather resource inequality is as follows:%
\begin{equation}
\left\langle \mathcal{N}\right\rangle +E\left[  qq\right]  \geq Q\left[
q\rightarrow q\right]  +R\left[  c\rightarrow c\right]  . \label{eq:GF}%
\end{equation}
The above resource inequality is again an asymptotic statement and its meaning
is similar to that in (\ref{eq:father}) and (\ref{eq:CQ}). The optimal rates
in the above resource inequality coincide with the father inequality
(\ref{eq:father}) when $R=0$, with the classical-quantum inequality
(\ref{eq:CQ})\ when $E=0$, and with the quantum capacity
\cite{PhysRevA.55.1613,capacity2002shor,ieee2005dev}\ when both $R=0$ and
$E=0$. The optimal strategy for the grandfather protocol is not time-sharing
the channel between father codes and entanglement-assisted classical codes. It
remains to be proven whether this optimal strategy outperforms time-sharing
\cite{prep2008dev}.

The goal of quantum error correction \cite{book2000mikeandike}\ is to find
efficient and practical ways of coding quantum information to protect it
against decoherence. One aspiration for this theory is to find quantum codes
that approach the rates given by quantum Shannon theory in the limit of large
block size.

The entanglement-assisted stabilizer formalism is a method for building a
quantum block code using entanglement \cite{science2006brun}. This theory has
several benefits, such as the ability to produce a quantum code from an
arbitrary classical linear block code, and has several generalizations
\cite{arx2007wilde,hsieh:062313,arx2007wildeEAQCC}. Entanglement-assisted
codes are \textquotedblleft father\textquotedblright\ codes, in the sense that
a good entanglement-assisted coding strategy should approach the optimal rates
in (\ref{eq:father}).

In this Proceeding, we design a framework for \textquotedblleft
grandfather\textquotedblright\ quantum codes. Our grandfather codes are useful
for the simultaneous transmission of classical and quantum information. Rather
than using block codes for this purpose, we design quantum convolutional
codes.\footnote{Kremsky, Hsieh, and Brun address the formulation of
grandfather block codes in a recent article \cite{prep2008hsieh}.} Quantum
convolutional coding is a recent extension of the stabilizer formalism
\cite{arxiv2004olliv,isit2006grassl,ieee2007forney}. One of the benefits of a
convolutional code is that it encodes a stream of information with an online
periodic encoding circuit. Our technique incorporates many of the known
techniques for quantum coding:\ subsystem codes
\cite{kribs:180501,poulin:230504}, entanglement-assisted codes
\cite{science2006brun}, convolutional codes
\cite{arxiv2004olliv,isit2006grassl,ieee2007forney}, and classical-quantum
coding \cite{cmp2005dev,beny:100502,prep2008hsieh}. The goal of our technique
is to provide a formalism for designing codes that approach the optimal triple
trade-off rates in the grandfather resource inequality in (\ref{eq:GF}). We
are not claiming that codes in our framework will reach capacity, but we are
instead providing a framework that one might later incorporate in a larger
theory, such as a quantum turbo coding theory \cite{arx2007poulin}.

We structure this proceeding as follows. Section~\ref{sec:hybrid} details our
\textquotedblleft grandfather\textquotedblright\ quantum convolutional codes.
We detail the finite-depth operations that quantum convolutional circuits
employ when encoding and decoding a stream of quantum information. We
explicitly show how to encode a stream of classical-quantum information using
finite-depth operations and discuss the error-correcting properties of our
codes. We end with an example of a grandfather quantum convolutional code. We
discuss errors that the code corrects actively and others that it corrects passively.

\section{Grandfather Quantum Convolutional Codes}

\label{sec:hybrid}We now detail the\ stabilizer formalism for our grandfather
quantum convolutional codes and describe how these codes operate. These codes are a significant
extension of the existing entanglement-assisted quantum convolutional codes
\cite{arx2007wildeEAQCC}.

An $\left[  n,k,l;r,c\right]  $ grandfather quantum convolutional code encodes
$k$ information qubits and $l$ information classical bits with the help of $c$
ebits, $a=n-k-l-c-r$ ancilla qubits, and $r$ gauge qubits. Each input frame
includes the following:

\begin{enumerate}
\item The sender Alice's half of $c$ ebits in the state $\left\vert \Phi
^{+}\right\rangle $.

\item $a=n-k-c-l-r$ ancilla qubits in the state $\left\vert 0\right\rangle $.

\item $r$ gauge qubits (which can be in any arbitrary state $\sigma$).

\item $l$ classical information bits $x^{1}\cdots x^{l}$, given by a
computational basis state $\left\vert x\right\rangle =X^{x^{1}}\otimes
\cdots\otimes X^{x^{l}}\left\vert 0\right\rangle ^{\otimes l}$.

\item $k$ information qubits in an arbitrary pure state $\left\vert
\psi\right\rangle $.\footnote{This statement is not entirely true because the
information qubits can be entangled across multiple frames, or with an
external system, but we use it to illustrate the idea.}
\end{enumerate}

The left side of Figure~\ref{fig:example}\ shows an example initial qubit
stream before an encoding circuit operates on it.

The stabilizer matrix $S_{0}\left(  D\right)  $\ for the initial qubit stream
is as follows:%
\begin{equation}
S_{0}\left(  D\right)  =\left[  \left.
\begin{array}
[c]{cccccc}%
I & I & 0 & 0 & 0 & 0\\
0 & 0 & 0 & 0 & 0 & 0\\
0 & 0 & I & 0 & 0 & 0
\end{array}
\right\vert
\begin{array}
[c]{cccccc}%
0 & 0 & 0 & 0 & 0 & 0\\
I & I & 0 & 0 & 0 & 0\\
0 & 0 & 0 & 0 & 0 & 0
\end{array}
\right]  , \label{eq:init-grand-stab}%
\end{equation}
where all identity matrices in the first two sets of rows are $c\times c$, the
identity matrix in the last row is $a\times a$, the three columns of all zeros
in both the \textquotedblleft Z\textquotedblright\ and \textquotedblleft
X\textquotedblright\ matrices are respectively $\left(  a+2c\right)  \times
r$, $\left(  a+2c\right)  \times l$, and $\left(  a+2c\right)  \times k$. The
matrices on the left of the vertical bar form the \textquotedblleft
Z\textquotedblright\ matrix and those on the right form the \textquotedblleft
X\textquotedblright\ matrix according to the Pauli-to-binary-polynomial
isomorphism (see Ref.'s~\cite{arx2007wilde,arx2007wildeEAQCC} for a review of
this isomorphism from the set of Pauli sequences to vectors of binary
polynomials). The first two sets of rows stabilize a set of $c$ ebits and the
last set of rows stabilizes a set of $a$ ancilla qubits. The first $c$ columns
of both the \textquotedblleft Z\textquotedblright\ and \textquotedblleft
X\textquotedblright\ matrix correspond to halves of ebits that the receiver
Bob possesses and the last $n$ columns in both matrices correspond to the
qubits that Alice possesses. The global generators (including both Alice and
Bob's qubits)\ form a commuting set of generators for all shifts, but Alice's
local generators do not necessarily form a commuting set.

Different sets of generators for the grandfather code are important in active
error correction, in passive error correction, and for the identification of
the $l$ classical information bits. We first write the unencoded generators
that act on the initial qubit stream. The first subgroup of generators is the
entanglement subgroup $\mathcal{S}_{E,0}$ with the following generators:%
\begin{equation}
S_{E,0}\left(  D\right)  =\left[  \left.
\begin{array}
[c]{ccccc}%
I & 0 & 0 & 0 & 0\\
0 & 0 & 0 & 0 & 0
\end{array}
\right\vert
\begin{array}
[c]{ccccc}%
0 & 0 & 0 & 0 & 0\\
I & 0 & 0 & 0 & 0
\end{array}
\right]  .\label{eq:grand-egroup}%
\end{equation}
The above generators are equivalent to the first two sets of rows in
(\ref{eq:init-grand-stab})\ acting on Alice's $n$ qubits. The next subgroup is
the isotropic subgroup $\mathcal{S}_{I,0}$ with the following generators:%
\begin{equation}
S_{I,0}\left(  D\right)  =\left[  \left.
\begin{array}
[c]{ccccc}%
0 & I & 0 & 0 & 0
\end{array}
\right\vert
\begin{array}
[c]{ccccc}%
0 & 0 & 0 & 0 & 0
\end{array}
\right]  .
\end{equation}
The above generators are equivalent to the last set of rows in
(\ref{eq:init-grand-stab})\ acting on Alice's $n$ qubits. The encoded versions
of both of the above two matrices are important for the active correction of
errors. The next subgroup is the gauge subgroup $\mathcal{S}_{G,0}$ whose
generators are as follows:%
\begin{equation}
S_{G,0}\left(  D\right)  =\left[  \left.
\begin{array}
[c]{ccccc}%
0 & 0 & I & 0 & 0\\
0 & 0 & 0 & 0 & 0
\end{array}
\right\vert
\begin{array}
[c]{ccccc}%
0 & 0 & 0 & 0 & 0\\
0 & 0 & I & 0 & 0
\end{array}
\right]  .
\end{equation}
The generators in $\mathcal{S}_{G,0}$ correspond to quantum operations that
have no effect on the encoded quantum information and therefore represent a
set of errors to which the code is immune. The last subgroup is the classical
subgroup $\mathcal{S}_{C,0}$ with generators%
\begin{equation}
S_{C,0}\left(  D\right)  =\left[  \left.
\begin{array}
[c]{ccccc}%
0 & 0 & 0 & I & 0
\end{array}
\right\vert
\begin{array}
[c]{ccccc}%
0 & 0 & 0 & 0 & 0
\end{array}
\right]  .\label{eq:grand-cgroup}%
\end{equation}
The grandfather code passively corrects errors corresponding to the encoded
version of the above generators because the initial qubit stream is immune to
the action of operators in $\mathcal{S}_{C,0}$ (up to a global phase). Alice
could measure the generators in $\mathcal{S}_{C,0}$ to determine the classical
information in each frame. Unlike quantum information, it is possible to
measure classical information without disturbing it.

Alice performs a periodic encoding circuit on her qubits to encode the initial
set of ebits, ancilla qubits, and information qubits in each frame. She
performs encoding operations only on her qubits because the channel spatially
separates her qubits from Bob's qubits. The periodic encoding circuit encodes
the information qubits and transforms the initial set of generators in
(\ref{eq:grand-egroup}-\ref{eq:grand-cgroup})\ to a more general set of
encoded generators. We use three types of operations in the example code in
Section~\ref{sec:example}:

\begin{enumerate}
\item Let $H\left(  i\right)  $ denote a Hadamard gate acting on qubit $i$ of
every frame. The effect of $H\left(  i\right)  $ is to swap column $i$ in the
\textquotedblleft Z\textquotedblright\ matrix with column $i$ in the
\textquotedblleft X\textquotedblright\ matrix.

\item Let $C\left(  i,j,D^{k}\right)  $ denote a CNOT\ gate from qubit $i$ in
every frame to qubit $j$ in a frame delayed by $k$ where $i\neq j$. This gate
affects both the \textquotedblleft X\textquotedblright\ and \textquotedblleft
Z\textquotedblright\ matrices. In the \textquotedblleft X\textquotedblright%
\ matrix, it multiplies column $i$ by $D^{k}$ and adds the result to column
$j$. In the \textquotedblleft Z\textquotedblright\ matrix, it multiplies
column $j$ by $D^{-k}$ and adds the result to column $i$. Let $f\left(
D\right)  $ be an arbitrary finite binary polynomial. Let $C\left(
i,j,f\left(  D\right)  \right)  $ denote the sequence of CNOT gates
corresponding to the polynomial $f\left(  D\right)  $.

\item Let $S\left(  i,j\right)  $ swap qubits $i$ and $j$ in every frame and
column $i$ and column $j$ in both \textquotedblleft X\textquotedblright\ and
\textquotedblleft Z.\textquotedblright
\end{enumerate}

Quantum convolutional circuits can employ other operations besides the above
three \cite{isit2006grassl}, but we need only these three operations for the
purposes of the current paper. The above operations and the others in the
above references are \textit{finite-depth}, because they transform any Pauli
sequence with a finite number of non-identity entries to a Pauli sequence with
a finite number of non-identity entries \cite{arx2007wildeEAQCC}. Finite-depth
operations are desirable because they do not propagate uncorrected errors into
the qubit stream when encoding or decoding.

The three gates used in this paper are all their own inverses. Therefore, the
operations of the decoding circuit are the encoding operations performed in
reverse order. The online nature of the decoding circuit follows directly from
the online nature of the encoding circuit.

Alice performs an encoding circuit with finite-depth operations to encode her
stream of qubits before sending them over the noisy quantum channel. The
encoding circuit transforms the initial stabilizer $S_{0}\left(  D\right)  $
to the encoded stabilizer $S\left(  D\right)  $\ as follows:%
\begin{equation}
S\left(  D\right)  =\left[  \left.
\begin{array}
[c]{cc}%
I & Z_{E1}\left(  D\right)  \\
0 & Z_{E2}\left(  D\right)  \\
0 & Z_{I}\left(  D\right)
\end{array}
\right\vert
\begin{array}
[c]{cc}%
0 & X_{E1}\left(  D\right)  \\
I & X_{E2}\left(  D\right)  \\
0 & X_{I}\left(  D\right)
\end{array}
\right]  ,\label{eq:grand-stab}%
\end{equation}
where $Z_{E1}\left(  D\right)  $, $X_{E1}\left(  D\right)  $, $Z_{E2}\left(
D\right)  $ and $X_{E2}\left(  D\right)  $ are each $c\times n$-dimensional,
and $Z_{I}\left(  D\right)  $ and $X_{I}\left(  D\right)  $ are both $a\times
n$-dimensional. The encoding circuit affects only the rightmost $n$ entries in
both the \textquotedblleft Z\textquotedblright\ and \textquotedblleft
X\textquotedblright\ matrix of $S_{0}\left(  D\right)  $ because these are the
qubits in Alice's possession. It transforms $S_{E,0}\left(  D\right)  $,
$S_{I,0}\left(  D\right)  $, $S_{G,0}\left(  D\right)  $, and $S_{C,0}\left(
D\right)  $ as follows:%
\begin{align}
S_{E}\left(  D\right)   &  =\left[  \left.
\begin{array}
[c]{c}%
Z_{E1}\left(  D\right)  \\
Z_{E2}\left(  D\right)
\end{array}
\right\vert
\begin{array}
[c]{c}%
X_{E1}\left(  D\right)  \\
X_{E2}\left(  D\right)
\end{array}
\right]  ,\\
S_{I}\left(  D\right)   &  =\left[  \left.
\begin{array}
[c]{c}%
Z_{I}\left(  D\right)
\end{array}
\right\vert
\begin{array}
[c]{c}%
X_{I}\left(  D\right)
\end{array}
\right]  ,\\
S_{G}\left(  D\right)   &  =\left[  \left.
\begin{array}
[c]{c}%
Z_{G1}\left(  D\right)  \\
Z_{G2}\left(  D\right)
\end{array}
\right\vert
\begin{array}
[c]{c}%
X_{G1}\left(  D\right)  \\
X_{G2}\left(  D\right)
\end{array}
\right]  ,\\
S_{C}\left(  D\right)   &  =\left[  \left.
\begin{array}
[c]{c}%
Z_{C}\left(  D\right)
\end{array}
\right\vert
\begin{array}
[c]{c}%
X_{C}\left(  D\right)
\end{array}
\right]  ,
\end{align}
where $Z_{G1}\left(  D\right)  $, $X_{G1}\left(  D\right)  $, $Z_{G2}\left(
D\right)  $ and $X_{G2}\left(  D\right)  $ are each $r\times n$-dimensional
and $Z_{C}\left(  D\right)  $ and $X_{C}\left(  D\right)  $ are each $l\times
n$-dimensional. The above polynomial matrices have the same commutation
relations as their corresponding unencoded polynomial matrices in
(\ref{eq:grand-egroup}-\ref{eq:grand-cgroup}) and respectively generate the
entanglement subgroup $\mathcal{S}_{E}$, the isotropic subgroup $\mathcal{S}%
_{I}$, the gauge subgroup $\mathcal{S}_{G}$, and the classical subgroup
$\mathcal{S}_{C}$.

The condition for a set of generators to form a commuting stabilizer is
equivalent to orthogonality of each row in $S\left(  D\right)  $ with respect
to the shifted symplectic product \cite{arxiv2004olliv,arx2007wilde}. This is
equivalent to the condition%
\begin{equation}
Z\left(  D^{-1}\right)  X^{T}\left(  D\right)  +X\left(  D^{-1}\right)
Z^{T}\left(  D\right)  =0,\label{eq:symp-ortho}%
\end{equation}
where $+$ represents binary addition of polynomials and the above matrix on
the right of the equality is an $\left(  n-k\right)  \times\left(  n-k\right)
$ null matrix. The original generators in (\ref{eq:init-grand-stab}) obey this
condition, and the periodic encoding circuit preserves the condition because
any encoding circuit preserves the commutation relations of the original generators.

A grandfather quantum convolutional code operates as follows. Alice begins
with an initial qubit stream as above. She performs the finite-depth encoding
operations corresponding to a specific grandfather quantum convolutional code.
She sends the encoded qubits online over the noisy quantum communication
channel. Bob combines the received qubits with his half of the ebits in each
frame. He obtains the error syndrome by measuring the generators in
(\ref{eq:grand-stab}). He processes these syndrome bits with a classical error
estimation algorithm to diagnose errors and applies recovery operations to
reverse the errors. He then performs the inverse of the encoding circuit to recover the initial
qubit stream with the information qubits and the classical information bits.
He recovers the classical information bits either by measuring the generators
in $\mathcal{S}_{C}$ before decoding or the generators in $\mathcal{S}_{C,0}$
after decoding.

A grandfather quantum convolutional code corrects errors in a Pauli error set
$\mathcal{E}$ that obey one of the following conditions $\forall E_{a}%
,E_{b}\in\mathcal{E}$:%
\[
\exists\ g\in\left\langle \mathcal{S}_{I},\mathcal{S}_{E}\right\rangle
:\left\{  g,E_{a}^{\dag}E_{b}\right\}  =0\text{ \ or \ }E_{a}^{\dag}E_{b}%
\in\left\langle \mathcal{S}_{I},\mathcal{S}_{G},\mathcal{S}_{C}\right\rangle
,
\]
where $\left\langle \cdot\right\rangle $ denotes the larger group generated by
a set of subgroups and $\left\{  A,B\right\}  $ denotes the anticomutator for
two operators $A$ and $B$ so that $\left\{  A,B\right\}  \equiv AB+BA$. It
corrects errors that anticommute with generators in $\left\langle
\mathcal{S}_{I},\mathcal{S}_{E}\right\rangle $ by employing a classical error
estimation algorithm, such as the Viterbi algorithm \cite{itit1967viterbi}.
The code passively protects against errors in the group $\left\langle
\mathcal{S}_{I},\mathcal{S}_{G},\mathcal{S}_{C}\right\rangle $.

Our scheme for quantum convolutional coding incorporates many of the known
techniques for quantum error correction. It can take full advantage of the
benefits of these different techniques.

\section{Example}

\label{sec:example}%
\begin{figure}
[ptb]
\begin{center}
\includegraphics[
natheight=9.979900in,
natwidth=11.080000in,
height=2.9698in,
width=3.2949in
]%
{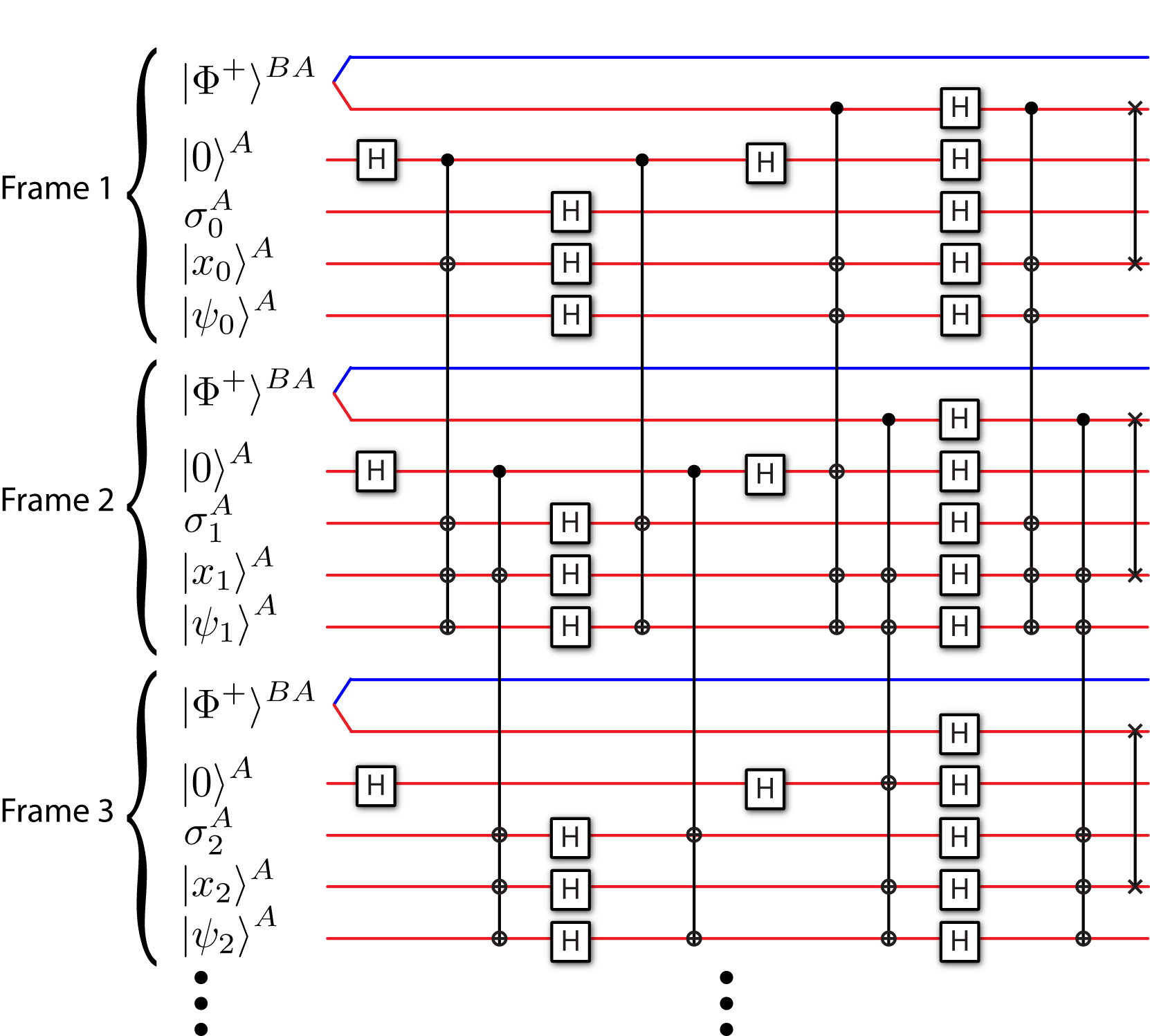}%
\caption{The encoding circuit for the grandfather quantum
convolutional code in our example. Each frame $i$\ has an ebit $\left\vert
\Phi^{+}\right\rangle ^{BA}$ shared between Bob and Alice, a fresh ancilla
qubit $\left\vert 0\right\rangle ^{A}$, a gauge qubit $\sigma_{i}^{A}$, an
information classical bit $\left\vert x_{i}\right\rangle ^{A}$, and an
information qubit $\left\vert \psi_{i}\right\rangle ^{A}$. Bob's qubits are
blue and Alice's qubits are red. Including the ebit and an ancilla qubit
implies that the code incorporates active quantum error correction. Including
the gauge qubit implies that the code has passive quantum error correction.
The code provides passive error correction of phase errors on the classical
bit. Alice encodes her classical bits and qubits using the encoding circuit
above. The decoding circuit consists of the above operations in reverse
order.}%
\label{fig:example}%
\end{center}
\end{figure}
We present an example of a grandfather quantum convolutional code in this
section. The code protects one information qubit and one classical bit with
the help of an ebit, an ancilla qubit, and a gauge qubit. The first frame of
input qubits has the state%
\begin{equation}
\rho_{0}=\left\vert \Phi^{+}\right\rangle \left\langle \Phi^{+}\right\vert
\otimes\left\vert 0\right\rangle \left\langle 0\right\vert \otimes\sigma
_{0}\otimes\left\vert x_{0}\right\rangle \left\langle x_{0}\right\vert
\otimes\left\vert \psi_{0}\right\rangle \left\langle \psi_{0}\right\vert ,
\end{equation}
where $\left\vert \Phi^{+}\right\rangle $ is the ebit, $\left\vert
0\right\rangle $ is the ancilla qubit, $\sigma_{0}$ is an arbitrary state for
the gauge qubit, $\left\vert x_{0}\right\rangle $ is a classical bit
represented by state $\left\vert 0\right\rangle $ or $\left\vert
1\right\rangle $, and $\left\vert \psi_{0}\right\rangle $ is one information
qubit equal to $\alpha_{0}\left\vert 0\right\rangle +\beta_{0}\left\vert
1\right\rangle $. The states of the other input frames have a similar form
though recall that information qubits can be entangled across multiple frames.

The initial unencoded stabilizer for the code is as follows:%
\[
S_{0}\left(  D\right)  =\left[  \left.
\begin{array}
[c]{cccccc}%
1 & 1 & 0 & 0 & 0 & 0\\
0 & 0 & 0 & 0 & 0 & 0\\
0 & 0 & 1 & 0 & 0 & 0
\end{array}
\right\vert
\begin{array}
[c]{cccccc}%
0 & 0 & 0 & 0 & 0 & 0\\
1 & 1 & 0 & 0 & 0 & 0\\
0 & 0 & 0 & 0 & 0 & 0
\end{array}
\right]  .
\]
The first two rows stabilize the ebit shared between Alice and Bob. Bob
possesses the half of the ebit in column one and Alice possesses the half of
the ebit in column two in both the left and right matrix. The third row
stabilizes the ancilla qubit.

The generators for the initial entanglement subgroup $\mathcal{S}_{E,0}$,
isotropic subgroup $\mathcal{S}_{I,0}$, gauge subgroup $\mathcal{S}_{G,0}$,
and classical subgroup $\mathcal{S}_{C,0}$ are respectively as follows:%
\begin{align*}
S_{E,0}\left(  D\right)   &  =\left[  \left.
\begin{array}
[c]{ccccc}%
1 & 0 & 0 & 0 & 0\\
0 & 0 & 0 & 0 & 0
\end{array}
\right\vert
\begin{array}
[c]{ccccc}%
0 & 0 & 0 & 0 & 0\\
1 & 0 & 0 & 0 & 0
\end{array}
\right]  ,\\
S_{I,0}\left(  D\right)   &  =\left[  \left.
\begin{array}
[c]{ccccc}%
0 & 1 & 0 & 0 & 0
\end{array}
\right\vert
\begin{array}
[c]{ccccc}%
0 & 0 & 0 & 0 & 0
\end{array}
\right]  ,\\
S_{G,0}\left(  D\right)   &  =\left[  \left.
\begin{array}
[c]{ccccc}%
0 & 0 & 1 & 0 & 0\\
0 & 0 & 0 & 0 & 0
\end{array}
\right\vert
\begin{array}
[c]{ccccc}%
0 & 0 & 0 & 0 & 0\\
0 & 0 & 1 & 0 & 0
\end{array}
\right]  ,\\
S_{C,0}\left(  D\right)   &  =\left[  \left.
\begin{array}
[c]{ccccc}%
0 & 0 & 0 & 1 & 0
\end{array}
\right\vert
\begin{array}
[c]{ccccc}%
0 & 0 & 0 & 0 & 0
\end{array}
\right]  .
\end{align*}
The sender performs the following finite-depth operations (order is from left
to right and top to bottom):%
\begin{align*}
&  H\left(  2\right)  \ C\left(  2,3,D\right)  \ C\left(  2,4,1+D\right)
\ C\left(  2,5,D\right)  \ H\left(  3,4,5\right)  \\
&  C\left(  2,3,D\right)  \ C\left(  2,5,D\right)  \ H\left(  2\right)
\ C\left(  1,2,D\right)  \ C\left(  1,4,1+D\right)  \\
&  C\left(  1,5,1+D\right)  \ H\left(  1,2,3,4,5\right)  \ C\left(
1,3,D\right)  \ C\left(  1,4,1+D\right)  \\
&  C\left(  1,5,1+D\right)  \ S\left(  1,4\right)  .
\end{align*}
Figure~\ref{fig:example}\ details these operations on the initial qubit
stream. The initial stabilizer matrix $S_{0}\left(  D\right)  $ transforms to
$S\left(  D\right)  =\left[  \left.
\begin{array}
[c]{c}%
Z\left(  D\right)
\end{array}
\right\vert
\begin{array}
[c]{c}%
X\left(  D\right)
\end{array}
\right]  $ under these encoding operations, where%
\begin{align}
Z\left(  D\right)   &  =\left[
\begin{array}
[c]{cccccc}%
1 & 0 & 0 & 0 & 0 & 0\\
0 & h\left(  D\right)   & D & 0 & 1 & h\left(  D\right)  \\
0 & 0 & 0 & D & D & D
\end{array}
\right]  ,\\
X\left(  D\right)   &  =\left[
\begin{array}
[c]{cccccc}%
0 & h\left(  D\right)   & 0 & D & 1 & h\left(  D\right)  \\
1 & 0 & 0 & 0 & 0 & 0\\
0 & 0 & 1 & 0 & 1 & 1
\end{array}
\right]  ,
\end{align}
and $h\left(  D\right)  =1+D$. The generators for the different subgroups
transform respectively as follows:%
\begin{multline*}
S_{E}\left(  D\right)  =\\
\left[  \left.
\begin{array}
[c]{ccccc}%
0 & 0 & 0 & 0 & 0\\
h\left(  D\right)   & D & 0 & 1 & h\left(  D\right)
\end{array}
\right\vert
\begin{array}
[c]{ccccc}%
h\left(  D\right)   & 0 & D & 1 & h\left(  D\right)  \\
0 & 0 & 0 & 0 & 0
\end{array}
\right]  ,
\end{multline*}%
\begin{align*}
S_{I}\left(  D\right)   &  =\left[  \left.
\begin{array}
[c]{ccccc}%
0 & 0 & D & D & D
\end{array}
\right\vert
\begin{array}
[c]{ccccc}%
0 & 1 & 0 & 1 & 1
\end{array}
\right]  ,\\
S_{G}\left(  D\right)   &  =\left[  \left.
\begin{array}
[c]{ccccc}%
0 & \frac{1}{D} & 1 & \frac{1}{D} & 0\\
0 & \frac{1}{D} & 0 & 0 & 0
\end{array}
\right\vert
\begin{array}
[c]{ccccc}%
0 & 0 & 0 & 0 & 0\\
0 & 0 & 1 & 0 & 0
\end{array}
\right]  ,
\end{align*}%
\begin{multline*}
S_{C}\left(  D\right)  =\\
\left[  \left.
\begin{array}
[c]{ccccc}%
1 & h\left(  \frac{1}{D}\right)   & 0 & h\left(  \frac{1}{D}\right)   & 0
\end{array}
\right\vert
\begin{array}
[c]{ccccc}%
0 & 0 & 0 & 0 & 0
\end{array}
\right]  .
\end{multline*}
The code actively protects against an arbitrary single-qubit error in every
other frame. One can check that the syndromes of the stabilizer in $S\left(
D\right)  $ satisfy this property. Consider the Pauli generators corresponding
to the generators in the entanglement subgroup and the isotropic subgroup:%
\begin{equation}
\cdots\left\vert
\begin{array}
[c]{ccccc}%
X & I & I & X & X\\
Z & I & I & Z & Z\\
I & X & I & X & X
\end{array}
\right\vert \left.
\begin{array}
[c]{ccccc}%
X & I & X & I & X\\
Z & Z & I & I & Z\\
I & I & Z & Z & Z
\end{array}
\right\vert \cdots,
\end{equation}
where all other entries in the left and right directions are tensor products
of the identity. We can use a table-lookup syndrome-based algorithm to
determine the error-correcting capability of the code. The method is similar
to the technique originally outlined in detail in Ref. \cite{ieee2007forney}.
The syndrome vector $s$\ consists of six bits where $s=s_{1}\cdots s_{6}$. The
first bit $s_{1}$ is one if the error anticommutes with the operator $XIIXX$
in the first part of the first generator above and zero otherwise. The second
bit $s_{2}$\ is one if the error anticommutes with the operator $XIXIX$ in the
delayed part of the first generator above and zero otherwise. The third
through sixth bits follow a similar pattern for the second and third
generators above. Table~\ref{TableKey}\ lists all single-qubit errors over
five qubits and their corresponding syndromes. The code corrects an arbitrary
single-qubit error in every other frame using this algorithm because the
syndromes are all unique. A syndrome-based Viterbi algorithm might achieve
better performance than the simple syndrome table-lookup algorithm outlined
above.%
\begin{table}[tbp] \centering
\begin{tabular}
[c]{l|l||l|l||l|l}\hline\hline
\textbf{Error} & \textbf{Syndrome} & \textbf{Error} & \textbf{Syndrome} &
\textbf{Error} & \textbf{Syndrome}\\\hline\hline
$X_{1}$ & 001100 & $X_{3}$ & 000001 & $X_{5}$ & 001101\\\hline
$Y_{1}$ & 111100 & $Y_{3}$ & 010001 & $Y_{5}$ & 111111\\\hline
$Z_{1}$ & 110000 & $Z_{3}$ & 010000 & $Z_{5}$ & 110010\\\hline
$X_{2}$ & 000100 & $X_{4}$ & 001001 &  & \\\hline
$Y_{2}$ & 000110 & $Y_{4}$ & 101011 &  & \\\hline
$Z_{2}$ & 000010 & $Z_{4}$ & 100010 &  & \\\hline\hline
\end{tabular}
\caption{A list of possible single-qubit errors in a particular frame and the corresponding
syndrome vector. The syndrome coresponding to any single-qubit error is unique. The code therefore
corrects an arbitrary single-qubit error in every other frame.}\label{TableKey}%
\end{table}%

This code also has passive protection against errors in $\left\langle
\mathcal{S}_{I},\mathcal{S}_{G},\mathcal{S}_{C}\right\rangle $. The Pauli form
of the errors in this group span over three frames and are as follows:%
\begin{equation}
\cdots\left\vert
\begin{array}
[c]{l}%
IIIII\\
IZIZI\\
IZIII\\
IZIZI
\end{array}
\right\vert
\begin{array}
[c]{l}%
IXIXX\\
IIZII\\
IIXII\\
ZZIZI
\end{array}
\left\vert
\begin{array}
[c]{l}%
IIZZZ\\
IIIII\\
IIIII\\
IIIII
\end{array}
\right\vert \cdots
\end{equation}
The smallest weight errors in this group have weight two and three. The code
passively corrects the above errors or any product of them or any
five-qubit shift of them.

There is a trade-off between passive error correction and the ability to
encode quantum information as discussed in Ref.~\cite{hsieh:062313}. One can
encode more quantum information by dropping the gauge group and instead
encoding an extra qubit. The gauge generators then become logical $X$ and $Z$
operators for the extra encoded qubits. One can also turn the classical bit
into a qubit by dropping the generators in the classical subgroup. These
generators then become logical $Z$ operators for the extra encoded qubits.

\section{Conclusion}

We have presented a framework and a representative example for grandfather
quantum convolutional codes. We have explicitly shown how these codes operate,
and how to encode and decode a classical-quantum information stream by using
ebits, ancilla qubits, and gauge qubits for redundancy. The ultimate goal for
this theory is to find quantum convolutional codes that might play an integral
part in larger quantum codes that approach the grandfather capacity
\cite{prep2008dev}. One useful line of investigation may be to combine this
theory with the recent quantum turbo-coding theory \cite{arx2007poulin}.

The authors thank Isaac Kremsky, Min-Hsiu Hsieh, and Igor Devetak for useful
discussions. MMW\ and TAB\ acknowledge support from NSF Grants CCF-0545845 and CCF-0448658.

\bibliographystyle{IEEEtran}
\bibliography{eaqcc-hybrid}

\end{document}